\documentclass[aps,prd,nofootinbib,twocolumn,groupedaddress]{revtex4}
\usepackage{epsfig}
\usepackage{amsfonts}
\usepackage{amssymb}
\usepackage{amsmath}
\usepackage{bm}
\usepackage{latexsym}
\usepackage{mathbbol}
\usepackage{graphicx}
\begin{document}
\title{Spin Filter in DVCS amplitudes}
\author{ Bernard L. G. Bakker$^{a}$ and Chueng-Ryong Ji$^{b}$\\
$^a$ Department of Physics and Astrophysics, Vrije Universiteit, 
     De~Boelelaan 1081, NL-1081 HV Amsterdam, 
     The Netherlands\\
$^b$ Department of Physics, 
North Carolina State University,
Raleigh, NC 27695-8202, 
USA}

\begin{abstract}
In deeply virtual Compton scattering (DVCS), it is found that in the
kinematics with large transverse photon momenta angular momentum is
not conserved if the amplitudes are calculated in terms of widely
used reduced operators.  Consequently, those kinematics will lead to
the wrong analysis of experimental data in terms of generalized parton
distributions. Moreover, the contribution of the longitudinally polarized
virtual photon in those kinematics should not be neglected in the analysis
of DVCS amplitudes.
\end{abstract}


\maketitle

For some time already, it has been realized that in non-forward
kinematics, e.g. deeply virtual Compton scattering (DVCS), the scattering
amplitudes, and thus cross sections, can be expressed in terms of objects,
generalized parton distributions (GPDs), which complement the knowledge
encoded in parton distribution functions \cite{ref.01,ref.02,ref.03}.
This idea has inspired many authors, whose work has been summarized in
several important review papers~\cite{ref.03-1,ref.03-2,ref.03-3}.

The paramount feature of the treatment of deep inelastic scattering (DIS)
and DVCS is factorization, i.e., writing the full scattering amplitude
as a convolution of a hard-scattering amplitude to be calculated in
perturbation theory, and a soft part embodying the hadronic structure. The
use of a hard photon that is far off-shell, say $-q^2 = Q^2 \gg$ any
relevant soft mass scale, enables factorization theorems~\cite{ref.04}
with the identification of the hard scattering amplitude.  A further step
is the introduction of light-front (LF) variables with the choice of a
preferred kinematics in which the amplitudes are calculated and the link
between the theoretical quantities, GPDs, and the cross sections can be
established. Light-front dynamics (LFD) (see e.g.  Ref.~\cite{ref.05})
can be invoked to further analyze the physics, as it has the advantage
that vacuum diagrams are either rigorously absent or suppressed. In the
context of DVCS it means that in a reference frame where the momentum
of the incoming photon $q^\mu$ has vanishing plus component: $q^+ \equiv
(q^0+ q^3)/\surd 2 = 0$, it cannot create partons, as their momenta must
have positive plus-components and these components are conserved in LFD.
This simplification facilitates the partonic interpretation of amplitudes.

This paper is devoted to a number of aspects of the derivation of
the relation of GPDs to the data that have been glossed over in
the literature.  We do so in the simplest possible setting, namely
DVCS on a structure-less spin-1/2 particle. Although this might seem
to preclude the discussion of the GPD formalism, we shall argue that
important lessons can be learnt from this exercise that are relevant to
the situation where factorization holds, the latter being essential for
a correct application of the GPD formalism.

Before we get into the discussion of the GPD formalism, we first report
our benchmark calculation of the complete full DVCS amplitude for the
scattering of a massless lepton $\ell$ off a point-like fermion $f$
of mass $m$.  In the final state, we find the scattered lepton $\ell'$,
the fermion $f'$ with momentum $k'$ and a (real) photon $\gamma'$, viz
$\ell\to\ell' + \gamma^*,\; \gamma^* + f \to \gamma' + f'$.  (`Complete'
means that the amplitude includes the leptonic part and `full' means that
no approximations are made in the calculation of the hadronic amplitude.)
The complete amplitude at tree level can be written as
\begin{equation}
 {\cal M} = \sum_h {\cal L}(\{\lambda',\lambda\} h) \frac{1}{q^2}
 {\cal H}(\{s',s\}\{h',h\}),
\label{eq.01}
\end{equation}
where the quantities $\lambda',\, \lambda,\, h',\, h,\, s'$, and $s$ are the
helicities of the outgoing and incoming leptons, outgoing and incoming
photons, and the rescattered and target fermions, respectively. 
Leaving out inessential factors, we may write
\begin{eqnarray}
 {\cal L}(\{\lambda',\lambda\} h) & = & \bar{u}(\ell';\lambda')
 {\epsilon\!\!/}^* (q;h) u(\ell;\lambda),
\nonumber \\
 {\cal H}(\{s',s\}\{h',h\})  & =  & \bar{u}(k';s')
 ({\cal O}_s + {\cal O}_u ) u(k;s), 
\label{eq.02}
\end{eqnarray}
where the $s$- and $u$-channel operators of the intermediate fermion are 
given by
\begin{eqnarray}
 {\cal O}_s &=&  \frac{{\epsilon\!\!/}^* (q';h')(k\!\!\!/
 + q\!\!\!/ + m)\epsilon\!\!/ (q;h)}{(k+q)^2 - m^2}, 
\nonumber \\
 {\cal O}_u &=&  \frac{\epsilon\!\!/ (q;h)(k\!\!\!/
 - {q\!\!\!/}^{\,\prime} + m){\epsilon\!\!/}^* (q';h')}{(k-q')^2 - m^2}.
\label{eq.03}
\end{eqnarray}

We take the following three kinematics for the momenta of the incoming
and outgoing particles in the hadronic amplitude:

(1) $\delta$-Kinematics ($q^+ \to 0$ as $\delta \to 0$)
\begin{eqnarray}
 q^\mu & = &
 \left(\delta p^+, Q,0, \frac{Q^2}{2(\zeta + \delta)p^+} + 
\frac{\zeta m^2}{2x(x-\zeta)p^+} \right),
 \nonumber \\
 {q'}^\mu & = &
 \left((\zeta + \delta)p^+, Q, 0, \frac{Q^2}{2(\zeta +\delta)p^+} \right),
 \nonumber \\
 k^\mu & = & \left(x p^+,0,0,\frac{m^2}{2xp^+} \right),
 \nonumber \\
 {k'}^\mu & = & \left((x-\zeta)p^+,0,0, \frac{m^2}{2(x-\zeta)p^+} \right),
\label{eq.04}
\end{eqnarray}

(2) $q'^+ = 0$ Kinematics (effectively, `1+1' dim.)
\begin{eqnarray}
 q^\mu & = &
 \left(-\zeta p^+, 0, 0, \frac{Q^2}{2\zeta p^+} \right),
 \nonumber \\
 {q'}^\mu & = &
 \left(0, 0, 0, \frac{Q^2}{2\zeta p^+} -
 \frac{\zeta m^2}{x(x-\zeta)p^+} \right).
\label{eq.05}
\end{eqnarray}
The momenta $k^\mu$ and $ {k'}^\mu$ are the same as in case (1).

(3) Nonvanishing $q^+$ and $q'^+$ Kinematics (with $m=0$)
\begin{eqnarray}
 q^\mu & = &
 \left(-\frac{\zeta}{2} p^+, \frac{Q}{\sqrt{2}},0, \frac{Q^2}{2\zeta p^+}
 \right),
 \nonumber \\
 {q'}^\mu & = &
 \left(\frac{\zeta}{2} p^+, \frac{Q}{\sqrt{2}},0, \frac{Q^2}{2\zeta p^+}
 \right).
\label{eq.06}
\end{eqnarray}
The momenta $k^\mu$ and $ {k'}^\mu$ are the same as in case (1) if
the limit $m \to 0$ is taken.  

These kinematics correspond to the
hard-scattering part of a DVCS amplitude where the fermions are the
quarks and $p^+$ is the plus-component of the momentum of the parent
hadron target. We use the Kogut-Soper spinors~\cite{ref.06} normalized to $2m$
and the polarization vectors
\begin{eqnarray}
 \epsilon(q;\pm 1) & = &
 \frac{1}{\surd 2} \left(0,\mp 1,-i,\mp \frac{q_x \pm i q_y}{q^+} \right),
 \nonumber \\
 \epsilon(q;0) & = &
 \frac{1}{\sqrt{q^2}} \left(q^+,q_x,q_y,\frac{q^2_\perp -q^2}{2q^+} \right),
\label{eq.07}
\end{eqnarray}
that correspond to the LF gauge $A^+ = 0$. 

All of these three kinematics yield identical kinematical invariants
such as $s = \frac{x-\zeta}{\zeta} Q^2$ and $u = - \frac{x}{\zeta} Q^2$
in the DVCS limit as $\delta \to 0$ and $m \to 0$. However, each of them
has its own merit of consideration.

In the $\delta \to 0$ limit, the $\delta$-kinematics coincides with
the well-known $q^+=0$ frame~\cite{ref.07} frequently cited in the
discussion of the GPD formalism. Noticing that taking $q^+=0$ will
lead to singular polarization vectors in the LF gauge $A^+ = 0$ (see
e.g. Eq.~(\ref{eq.07})), we proceed with care: $q^+$ is set to $\delta
p^+$ and all amplitudes are expanded in powers of $\delta$, taking the
limit $\delta\to 0$ at the very end of the calculation of the complete,
physical amplitude.  The $q'^+ =0$ kinematics without any transverse
component (effectively, `1+1' dimensional) avoids the singularity in
the polarization vectors of the real photon and consequently provides
a convenient framework of calculation without encountering any
singularity. Similarly, the nonvanishing $q^+$ and $q'^+$ kinematics
also avoids the singularity in the amplitude calculation, while the
photons carry the same order of transverse momenta as the ones in the
$\delta$-kinematics given by Eq.~(\ref{eq.04}).

\begin{table}
\caption{\label{tab.01} Leptonic amplitudes in kinematics 
corresponding to Eqs.~(\protect\ref{eq.04})-(\protect\ref{eq.06})}
\begin{tabular}{|rl|r|c|c|}
\hline 
 &     & \multicolumn{3}{c|}{${\cal L} (\{\lambda',\lambda\}h)$} \\
\hline
 $\{\lambda',\,\lambda\}$ & $h$ & Eq.~(4) \quad & Eq.~(5) & Eq.~(6) \\
\hline 
 $\{\frac{1}{2}, \frac{1}{2}\}$ & $+1$ &
 $ -Q \left(1 - \frac{\delta}{4\zeta} + \frac{2 \zeta}{\delta}\right)$ &
 0 & $2 Q$ \\
 $\{\frac{1}{2}, \frac{1}{2}\}$ & $-1$ &
 $- Q \left(1 - \frac{3\delta}{4\zeta} - \frac{2 \zeta}{\delta}\right)$ &
 $-2 Q$ & $-4 Q$ \\
 $\{\frac{1}{2},\frac{1}{2}\}$ & $~0$ & $-i 2\surd 2 Q\, \frac{\zeta}{\delta}$ &
 0 & $4 i Q$ \\
\hline
\end{tabular}
\end{table}

\begin{figure}
\epsfig{file=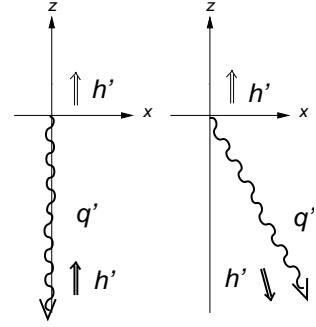,width=40mm} \hspace{5mm}
\caption{Spin directions corresponding to an LF boost in the $z$-direction
only, l.h.s, and one including transverse parts, r.h.s., from a state with
initial spin in the $+z$-direction. Note that the spin does not align
completely in the latter case.}
\label{fig.02}
\end{figure}

The results from these three kinematics are summarized in Tables~\ref{tab.01},
\ref{tab.02}, and \ref{tab.03}.  A straightforward evaluation of
${\cal L} (\{\lambda',\lambda\}h)$ gives the result in Table~I, where
we have used the corresponding lepton kinematics\footnote{The details
of lepton kinematics and spinors will be presented somewhere else.}
to Eqs.~(\ref{eq.04})-(\ref{eq.06}) and presented the results only
up to order $\delta$ as well as in the DVCS limit.  For the massless
leptons helicity is conserved. The amplitudes not shown in Table~I can
be obtained using the helicity rule
\begin{equation}
 {\cal L}(\{-\lambda', -\lambda\}~-h) =
 (-1)^{\lambda'-\lambda +h}  {\cal L}(\{\lambda', \lambda\}~h) .
\label{eq.08}
\end{equation}
The full hadronic amplitudes are shown in Table~\ref{tab.02}, where we again
presented the results only up to order $\delta$. They
obey the rule
\begin{equation}
 {\cal H}(\{-h', -h\}\{-s',-s\}) = (-1)^{h-h'-s+s'}{\cal H}(\{h', h\}\{s',s\}) .
\label{eq.09}
\end{equation}
\begin{table}
\caption{\label{tab.02} Hadronic amplitudes in DVCS  in three kinematics
given by Eqs.~(\protect\ref{eq.04})-(\protect\ref{eq.06})}
\begin{tabular}{|cc|l|c|c|}
\hline 
 &     & \multicolumn{3}{c|}{${\cal H} (\{h',h\} \{s',s\})$} \\
\hline
 $\{h',\,h\}$ & $\{s',s\}$ & \quad Eq.~(4) & Eq.~(5) & Eq.~(6) \\
\hline 
 ~$\{+1, +1\}$ & $\{\frac{1}{2},\frac{1}{2}\}$~ &
 ~$ 2\sqrt{\frac{x}{x-\zeta}} \, \left(1 + \frac{ \zeta}{\delta}\right)$  &
 $2 \sqrt{\frac{x-\zeta}{x}}$ &  $-2\sqrt{\frac{x}{x-\zeta}}$ \\
 ~$\{+1, +1\}$ & $\{-\frac{1}{2},-\frac{1}{2}\}$~ &
 ~$ 2\sqrt{\frac{x-\zeta}{x}} \, \left(1 + \frac{ \zeta}{\delta}\right)$  &
 $2 \sqrt{\frac{x}{x-\zeta}}$ &  $-2\sqrt{\frac{x-\zeta}{x}}$ \\
 ~$\{+1, -1\}$ & $\{\frac{1}{2},\frac{1}{2}\}$~ &
 ~$ - 2\sqrt{\frac{x}{x-\zeta}}\,  \frac{\zeta}{\delta}$  &
 $0$ &  $4\sqrt{\frac{x}{x-\zeta}}$ \\
 ~$\{+1, -1\}$ & $\{-\frac{1}{2},-\frac{1}{2}\}$~ &
 ~$ - 2\sqrt{\frac{x-\zeta}{x}}\,  \frac{ \zeta}{\delta}$  &
 $0$ &  $4\sqrt{\frac{x-\zeta}{x}}$ \\
 ~$\{+1, ~0\}$ & $\{\frac{1}{2},\frac{1}{2}\}$~ &
 ~$ i \sqrt{2}\sqrt{\frac{x}{x-\zeta}}
 \, \left(1 + \frac{2\zeta}{\delta}-\frac{\delta}{4\zeta} \right)$  &
 $0$ & $-i 4\sqrt{\frac{x}{x-\zeta}}$ \\
 ~$\{+1, ~0\}$ & $\{-\frac{1}{2},-\frac{1}{2}\}$~ &
 ~$ i \sqrt{2}\sqrt{\frac{x-\zeta}{x}}
 \, \left(1 + \frac{2\zeta}{\delta}-\frac{\delta}{4\zeta} \right)$  &
 $0$ &  $-i 4\sqrt{\frac{x-\zeta}{x}}$ \\
\hline
\end{tabular}
\end{table}
The complete DVCS amplitude ${\cal M}$ in Eq.~(\ref{eq.01}) is shown in
Table~\ref{tab.03}. Since all the singular terms of orders $\delta^{-2}$
and $\delta^{-1}$ are exactly cancelled out in the complete amplitude,
we have taken $\delta=0$ in Table~\ref{tab.03}.
\begin{table*}
\caption{\label{tab.03} Complete DVCS amplitudes, in three kinematics
given by Eqs.~(\protect\ref{eq.04})-(\protect\ref{eq.06})}
\begin{tabular}{|ccc|c|c|c|}
\hline
 &  &  & \multicolumn{3}{c|}{$\sum_h {\cal L}(\{\lambda',\lambda\}, h)\frac{1}{q^2}{\cal H}
 (\{h',h\} \{s',s\})$} \\
\hline
 $\{\lambda',\,\lambda\}$ & $h'$ & $\{s',s\}$ & Eq.~(4) & Eq.~(5) & Eq.~(6)\\
\hline
 ~$\{\frac{1}{2}, \frac{1}{2}\}$ & $1$ & $\{\frac{1}{2},\frac{1}{2}\}$~ &
 ~$ \frac{4}{Q}\sqrt{\frac{x}{x-\zeta}}$  & $0$ &  $\frac{4}{Q}\sqrt{\frac{x}{x-\zeta}}$ \\
 ~$\{\frac{1}{2}, \frac{1}{2}\}$ & $1$ & $\{-\frac{1}{2},-\frac{1}{2}\}$ &
 ~$ \frac{4}{Q}\sqrt{\frac{x-\zeta}{x}}$  & $0$ &  $\frac{4}{Q}\sqrt{\frac{x-\zeta}{x}}$ \\
 ~$\{-\frac{1}{2}, -\frac{1}{2}\}$ & $1$ & $\{\frac{1}{2},\frac{1}{2}\}$~ &
 ~$0$ & $-\frac{4}{Q}\sqrt{\frac{x-\zeta}{x}}$ &  $0$ \\
 ~$\{-\frac{1}{2}, -\frac{1}{2}\}$ & $1$ & $\{-\frac{1}{2},-\frac{1}{2}\}$ &
 ~$0$ & $-\frac{4}{Q}\sqrt{\frac{x}{x-\zeta}}$ & $0$ \\
\hline
 ~$\{\frac{1}{2}, \frac{1}{2}\}$ & $-1$ & $\{\frac{1}{2},\frac{1}{2}\}$~ &
 ~$0$ & $ \frac{4}{Q}\sqrt{\frac{x}{x-\zeta}}$  & $0$  \\
 ~$\{\frac{1}{2}, \frac{1}{2}\}$ & $-1$ & $\{-\frac{1}{2},-\frac{1}{2}\}$ &
 ~$0$ & $\frac{4}{Q}\sqrt{\frac{x-\zeta}{x}}$  & $0$ \\
 ~$\{-\frac{1}{2}, -\frac{1}{2}\}$ & $-1$ & $\{\frac{1}{2},\frac{1}{2}\}$~ &
 ~$-\frac{4}{Q}\sqrt{\frac{x-\zeta}{x}}$ &  $0$ & $-\frac{4}{Q}\sqrt{\frac{x-\zeta}{x}}$ \\
 ~$\{-\frac{1}{2}, -\frac{1}{2}\}$ & $-1$ & $\{-\frac{1}{2},-\frac{1}{2}\}$ &
 ~$-\frac{4}{Q}\sqrt{\frac{x}{x-\zeta}}$ & $0$ & $-\frac{4}{Q}\sqrt{\frac{x}{x-\zeta}}$ \\
\hline
\end{tabular}
\end{table*}
Note in Table~\ref{tab.03} that there is an interchange\footnote{We
have also confirmed the similar interchange of the helicity amplitudes
between the kinematics with and without the transverse momentum of the
virtual photon in the case of a form-factor calculation.}
of the polarization of the final photon in the result of the `1+1'
dim. kinematics in comparison with the other kinematics, in which the
momenta of photons have transverse components. This is remarkable in
view of the LF helicity~\cite{ref.08}.  To appreciate this point, we
draw in Fig.~\ref{fig.02} the spin directions of the outgoing photon
with the LF helicity $h'$ for the two different kinematics: one without
any transverse momentum such as Eq.~(\ref{eq.05}) and the other
with the transverse momentum of order $Q$ such as Eq.~(\ref{eq.04})
or Eq.~(\ref{eq.06}). One should realize that the LF helicity states
are defined for a momentum $q'$ by taking a state at rest with the spin
projection along the $z$ direction equal to the desired helicity, then
boosting in the $z$ direction to get the desired $q'^+$, and then doing
a LF transverse boost (i.e., $E_1 = K_1 + J_2$ \cite{ref.08}) to get the
desired transverse momentum $\vec{q'}_\perp$.  Whether the kinematics
includes the LF transverse boost ($E_1$) or not makes a dramatic
difference in the spin direction because $E_1$ rotates the spin direction.
Thus, for the l. h. panel of Fig.~\ref{fig.02}, the spin direction of
the LF helicity state is opposite (or antiparallel) to the direction
of the photon momentum while for the r. h. panel of Fig.~\ref{fig.02},
the spin directions of the LF helicity state and the Jacob-Wick helicity
state~\cite{ref.09} are related~\cite{ref.08} by the Wigner function
$d^1_{h',h'}(\tan^{-1}\frac{2m}{Q})$ in the DVCS limit, which becomes
unity as $Q \to \infty$. This illustrates the correspondence between
the results of a kinematics with $\vec{q'}_\perp = 0$ and a kinematics
with the transverse momentum of order $Q$: e.g. in Table~\ref{tab.03},
the result of $h'=1$ in the effective `1+1D' kinematics corresponds to the
result of $h'=-1$ in the $\delta$-kinematics or the nonvanishing $q^+$ and
$q'^+$ kinematics for ${\lambda',\lambda}={\frac{1}{2},\frac{1}{2}}$ and
${s',s}={\frac{1}{2},\frac{1}{2}}$.  One should note that the conservation
of angular momentum is satisfied in the complete full amplitudes for
\underline{any} kinematics. Therefore, we may take the calculation up to
now as a benchmark for the discussion of the GPD formalism as we do below.
Rewriting the $s$- and $u$- channel hadronic amplitudes as
\begin{eqnarray}
\bar{u}(k';s') {\cal O}_s u(k;s) &=& 
 {\epsilon_\mu}^* (q';h')\epsilon_\nu (q;h) {T_s}^{\mu\nu}, 
\nonumber \\
\bar{u}(k';s') {\cal O}_u u(k;s) &=& 
 {\epsilon_\mu}^* (q';h')\epsilon_\nu (q;h) {T_u}^{\mu\nu}, 
\label{eq.10}
\end{eqnarray}
we may neglect an inessential fermion mass $m$ to express the tensorial
amplitudes ${T_s}^{\mu\nu}$ and ${T_u}^{\mu\nu}$ as
\begin{eqnarray}
 {T_s}^{\mu\nu} &=& \frac{k_\alpha +q_\alpha}{s}
 \bar{u}(k';s') \gamma^\mu \gamma^\alpha \gamma^\nu u(k;s), 
\nonumber \\
 {T_u}^{\mu\nu} &=& \frac{k_\alpha -q_\alpha}{u}
 \bar{u}(k';s') \gamma^\nu \gamma^\alpha \gamma^\mu u(k;s), 
\label{eq.11}
\end{eqnarray}
respectively. Using the identity 
\begin{equation}
 \gamma^\mu \gamma^\alpha \gamma^\nu = g^{\mu\alpha}\gamma^\nu
 + g^{\alpha\nu}\gamma^\mu-g^{\mu\nu}\gamma^\alpha
 + i\epsilon^{\mu\alpha\nu\beta}\gamma_\beta \gamma_5
\label{eq.12}
\end{equation}
and the Sudakov variables $n^\mu(+)=(1,0,0,0)$ and $n^\mu(-)=(0,0,0,1)$,
one may expand ${T_s}^{\mu\nu}$ and ${T_u}^{\mu\nu}$ to find the terms
proportional to $\bar{u}(k';s') n\!\!\!/ (-) u(k;s)$ and $\bar{u}(k';s')
n\!\!\!/ (-) \gamma_5 u(k;s)$ that correspond to the nucleon GPDs
$H(x,\Delta^2,\zeta)$ and ${\bar H}(x,\Delta^2,\zeta)$ defined e.g. in
Ref.~\cite{ref.01}, respectively (here, $\Delta^2 = (q'-q)^2$).  One should
note, however, that a special system of coordinates without involving
any large transverse momentum (see e.g. Eq.~(\ref{eq.05})) was chosen in
Ref.~\cite{ref.01} to compute the scattering amplitude in terms of GPDs.
We realize that Ref.~\cite{ref.02} uses essentially the same special
system of coordinates as Ref.~\cite{ref.01}.

In order to cover the more general kinematics involving large transverse
momenta such as given in Eqs.~(\ref{eq.04}) and (\ref{eq.06}), we
may expand $q^\mu$ (similarly $q'^\mu$) and $k^\mu$ as $q^\mu = q^+
n^\mu(+) +q^- n^\mu(-) +{q_\perp}^\mu$ and $k^\mu = k^+ n^\mu(+) +k^-
n^\mu(-)$ with ${q_\perp}^\mu$ representing the transverse momentum
corresponding to $q^\mu$. For $m=0$, $k^-=0$ and ${T_s}^{\mu\nu}$
(similarly ${T_u}^{\mu\nu}$) can be expanded as
\begin{eqnarray}
&& {T_s}^{\mu\nu} = \frac{1}{s}
 \left[
       \left(
\{(k^+ +q^+ )n^\mu (+) +q^- n^\mu (-) +{q_\perp}^\mu \}n^\nu (+)
\right.\right.
\nonumber\\
&& \left.\left.
+ \{(k^+ +q^+ )n^\nu (+) +q^- n^\nu (-) +{q_\perp}^\nu \}n^\mu (+)
- g^{\mu\nu} q^-
        \right)
 \right.
\nonumber \\
&&
 \left.
 \times  \bar{u}(k';s') n\!\!\!/ (-) u(k;s)
 \right.
\nonumber \\
&& \left.
-i\epsilon^{\mu\nu\alpha\beta}
\{(k^+ +q^+ )n_\alpha (+) +q^- n_\alpha (-) +{q_\perp}_\alpha \}
 n_\beta (+)
 \right. 
\nonumber \\
&& 
 \left.
 \times \bar{u}(k';s') n\!\!\!/ (-) \gamma_5 u(k;s)
 \right].
\label{eq.13}
\end{eqnarray}
Since $q^-$ has the highest power of $Q$ among the components of
momenta, one may just take the terms proportional to $q^-$ as shown in
Refs.~\cite{ref.01} and \cite{ref.02}, i.e.,
\begin{eqnarray}
 {T_s}^{\mu\nu}  & = &
 \frac{q^-}{s}\left[\{n^\mu (-) n^\nu (+) + n^\nu (-) n^\mu (+)
- g^{\mu\nu}\}
 \right.
\nonumber \\
 & &
 \left.
 \times \bar{u}(k';s') n\!\!\!/ (-) u(k;s)
 \right.
\nonumber \\
 & &
 \left.
-i\epsilon^{\mu\nu\alpha\beta}n_\alpha (-) n_\beta (+)
 \times \bar{u}(k';s') n\!\!\!/ (-) \gamma_5 u(k;s)\right].
\nonumber \\
\label{eq.14}
\end{eqnarray}
However, one should note that Eq.~(\ref{eq.14}) cannot provide the full
result of the hadronic amplitude in the kinematics involving large
transverse momenta such as Eq.~(\ref{eq.04}) and Eq.~(\ref{eq.06}),
because the polarization vectors ${\epsilon_\mu}^* (q';h')$ and
$\epsilon_\nu (q;h)$ in Eq.~(\ref{eq.10}) amplify the contributions
neglected in the tensorial amplitude ${T_s}^{\mu\nu}$ (similarly
${T_u}^{\mu\nu}$) given by Eq.~(\ref{eq.14}).  For example, the coefficient
of $\bar{u}(k';s') n\!\!\!/ (-) u(k;s)$ in the $s$-channel hadronic
amplitude $\bar{u}(k';s') {\cal O}_s u(k;s)$ is given by the following
four terms:
\begin{eqnarray}
 & \frac{1}{s} &\left[\,
 2(k^+ + q^+) \epsilon^{*-} (q';h') \epsilon^- (q;h) 
 \right.
\nonumber \\
 &  & +
 \left.
 \epsilon^{*-} (q';h')\; q_\perp \cdot \epsilon_\perp (q;h) 
 \right.
\nonumber \\
 & & +
 \left.
  \epsilon^{-} (q;h)\; q_\perp \cdot {\epsilon_\perp}^* (q';h')
 \right.
\nonumber \\
 & & -
 \left.
 q^- {\epsilon_\perp}^* (q';h') \cdot \epsilon_\perp (q;h)\,\right].
\label{eq.15}
\end{eqnarray}
Since all of the above four terms have the same powers of $Q$, one
must keep them all.  In other words, the
factorization in the tensorial amplitude ${T_s}^{\mu\nu} + {T_u}^{\mu\nu}$
cannot hold in general because the polarization vectors ${\epsilon_\mu}^*
(q';h')$ and $\epsilon_\nu (q;h)$ can amplify the terms neglected in the
tensorial amplitude unless a special system of coordinates is chosen to
avoid the large transverse momenta of initial and final photons such
as given by Eq.~(\ref{eq.05}).  Thus, we note that the formulation
of GPDs on the level of the tensorial amplitude is not general enough
to cover the kinematics with large transverse momenta such as given by
Eqs.~(\ref{eq.04}) and (\ref{eq.06}) but is limited to the special
system of coordinates without involving large transverse momenta as
given by Eq.~(\ref{eq.05}).  This is the main point of this paper. In
the following, we demonstrate this point explicitly, presenting
the consequence of taking the reduced amplitude that keeps only the
terms proportional to $q^-$ in the tensorial amplitude as done in the
formulation of GPDs. Unless the kinematics is chosen properly to avoid
the large transverse momenta of initial and final photons, we find that
the reduced amplitude does not agree with the full amplitude but yield the
wrong result, not even satisfying the conservation of angular momentum.

Since the $q^+ = 0$ frame is used~\cite{ref.07} in the GPD formalism,
we utilize the $\delta$-kinematics for our demonstration. We perform
an expansion in the hard momentum scale $Q$, which allows us to define
reduced hadronic amplitudes.  In the expansion, it is important to
retain terms of orders $\delta^{-1}$, \dots $\delta^2$ as well as
orders $Q^{-1}$, \dots $Q^2$, as it turns out that not only are the
order $\delta^{-1}$-terms cancelled by order $\delta$ terms in the
convolution of ${\cal L}$ and ${\cal H}$, but also that the order
$Q^{-1}$-contribution of the longitudinally polarized virtual photon
gives a finite contribution in leading order. (We have checked that the
two limits, $\delta\to 0$ and $Q\to\infty$ commute.)

The reduced hadronic operators used in the formulation of GPDs are defined
as the limits $Q\to\infty$ of the operators given in Eq.~(\ref{eq.03})
and found to be, as expected:
\begin{eqnarray}
 {\cal O}_s|_{\rm Red} &=&
 \frac{{\epsilon\!\!/}^* (q';h')\gamma^+\epsilon\!\!/ (q;h)}{2 p^+} \,
 \frac{1}{x-\zeta}, 
\nonumber \\
 {\cal O}_u|_{\rm Red} &=&
 \frac{\epsilon\!\!/ (q;h)\gamma^+{\epsilon\!\!/}^* (q';h')}{2 p^+} \, 
 \frac{1}{x}.
\label{eq.16}
\end{eqnarray}
These reduced propagators contain the nilpotent Dirac matrix $\gamma^+$
only, which kills the singular parts of the polarization vectors, namely
$\epsilon^-(q;h) \gamma^+$. This is the  reason for disregarding the
singularities in the polarization vectors in $q^+ = 0$ kinematics, as the
reduced hadronic amplitude does not `see' it. However, the leptonic part
${\cal L}$ of the complete amplitude is also singular.  Consequently,
the complete amplitude calculated with the reduced hadronic part and
taking into account the transverse polarizations only, is wrong, even
in the limit $Q\to\infty$.

\begin{table*}[ht]
\caption{\label{tab.04} Complete amplitudes in $\delta$-kinematics}
\begin{tabular}{|ccc|l|l|}
\hline 
 ~$\{\lambda',\lambda\}$ & ~$\{h', h\}$ & ~ $\{s',s\}$
 &~ ${\cal L} \frac{1}{q^2}{\cal H}_{\rm Full}$ 
 &~ ${\cal L} \frac{1}{q^2}{\cal H}_{\rm Red}$  \\
\hline
~$\{\frac{1}{2},\frac{1}{2}\}$ & ~$\{+1, +1\}$ & ~ $\{\frac{1}{2},\frac{1}{2}\}$
 &~ $\frac{1}{Q}\sqrt{\frac{x}{x-\zeta}}\,
 \left(\frac{4\zeta^2}{\delta^2}+\frac{6\zeta}{\delta} +\frac{3}{2}
 -\frac{\delta}{4\zeta} \right)$ &
 ~ $\frac{2}{Q}\sqrt{\frac{x-\zeta}{x}}\,
 \left(\frac{2\zeta}{\delta} +1 -\frac{\delta}{4\zeta} \right)$  \\
~$\{\frac{1}{2},\frac{1}{2}\}$ & ~$\{+1, ~0\}$ & ~ $\{\frac{1}{2},\frac{1}{2}\}$
 &~ $\frac{1}{Q}\sqrt{\frac{x}{x-\zeta}}\,
 \left(\frac{-8\zeta^2}{\delta^2}-\frac{4\zeta}{\delta} +1
 -\frac{\delta}{2\zeta} \right)$  &
 ~ $\frac{2}{Q}\sqrt{\frac{x-\zeta}{x}}\,
 \left(-\frac{2\zeta}{\delta} -1 +\frac{\delta}{4\zeta} \right)$  \\
~$\{\frac{1}{2},\frac{1}{2}\}$ & ~$\{+1, -1\}$ & ~ $\{\frac{1}{2},\frac{1}{2}\}$
 &~ $\frac{1}{Q}\sqrt{\frac{x}{x-\zeta}}\,
 \left(\frac{4\zeta^2}{\delta^2}-\frac{2\zeta}{\delta} +\frac{3}{2}
 -\frac{5\delta}{4\zeta} \right)$ &~ $0$ \\
\hline
 ~ & ~$\sum_h$ & ~ &~
 $\frac{1}{Q}\sqrt{\frac{x}{x-\zeta}}\,\left( 4 - \frac{2\delta}{\zeta} \right)$ &~ $0$\\
\hline
 ~$\{\lambda',\lambda\}$ & ~$\{h', h\}$ & ~ $\{s',s\}$
 &~ ${\cal L} \frac{1}{q^2}{\cal H}_{\rm Full}$ & ~ ${\cal L} \frac{1}{q^2}{\cal H}_{\rm Red}$ \\
\hline 
~$\{\frac{1}{2},\frac{1}{2}\}$ & ~$\{-1, +1\}$ & ~ $\{\frac{1}{2},\frac{1}{2}\}$
 &~ $\frac{1}{Q}\sqrt{\frac{x-\zeta}{x}}\,
 \left(-\frac{4\zeta^2}{\delta^2}-\frac{2\zeta}{\delta} +\frac{1}{2}
 -\frac{\delta}{4\zeta} \right)$ &~ $0$ \\
~$\{\frac{1}{2},\frac{1}{2}\}$ & ~$\{-1, ~0\}$ & ~ $\{\frac{1}{2},\frac{1}{2}\}$
 &~ $\frac{1}{Q}\sqrt{\frac{x-\zeta}{x}}\,
 \left(\frac{8\zeta^2}{\delta^2}+\frac{4\zeta}{\delta} -1
 +\frac{\delta}{2\zeta} \right)$
 &~ $\frac{2}{Q}\sqrt{\frac{x}{x-\zeta}}\,
 \left(\frac{2\zeta}{\delta} +1 -\frac{\delta}{4\zeta} \right)$  \\
~$\{\frac{1}{2},\frac{1}{2}\}$ & ~$\{-1, -1\}$ & ~ $\{\frac{1}{2},\frac{1}{2}\}$
 &~ $\frac{1}{Q}\sqrt{\frac{x-\zeta}{x}}\,
 \left(-\frac{4\zeta^2}{\delta^2}-\frac{2\zeta}{\delta} +\frac{1}{2}
 -\frac{\delta}{4\zeta} \right)$
 &~ $\frac{2}{Q}\sqrt{\frac{x}{x-\zeta}}\,
 \left(-\frac{2\zeta}{\delta} +1 -\frac{3\delta}{4\zeta} \right)$  \\
\hline
 ~ & ~$\sum_h$ & ~ &~ 0 
 &~ $\frac{1}{Q}\sqrt{\frac{x}{x-\zeta}}\,\left( 4 - \frac{2\delta}{\zeta} \right)$\\
\hline
\end{tabular}
\end{table*}
Table~\ref{tab.04} clearly shows that the reduced amplitudes and the full
ones disagree.  We have checked that the same disagreement occurs in the
nonvanishing $q^+$ and $q'^+$ kinematics given by Eq.~(\ref{eq.06}),
although for the kinematics without any transverse component,
e.g. Eq.~(\ref{eq.05}), the reduced amplitudes and the full ones do
agree.  Upon convoluting the leptonic and hadronic amplitudes to obtain
the complete ones, we find that the singular $1/\delta$-terms cancel in
$\delta$-kinematics, but the full and reduced hadronic amplitudes do not
produce the same complete ones.  Moreover, if the contribution of the
longitudinal polarization of the virtual photon is neglected, i.e., if
its propagator is reduced too, the singular parts do not cancel out. So,
the contribution of the longitudinal part, contrary to expectations,
is not suppressed by a factor $1/Q$ compared to the contributions of
the transversely polarized photons.  As such, the contribution of the
longitudinal polarization should not be neglected in the kinematics
given by Eqs.~(\ref{eq.04}) and (\ref{eq.06}), where the photons
carry transverse momenta of order $Q$.

We see here that summing the complete amplitudes over $h$ gives the
same result for the full and the reduced amplitudes, {\it but for
the interchange of the polarization of the final photon}. As this
polarization is an observable, we observe that the reduced amplitude
gives the wrong amplitude.  Clearly the tree-level hard amplitude plays
the role of a spin filter. Using the reduced amplitudes means using a
spin filter that provides an erroneous connection between the data and
the GPD.  Using it only for the spin-averaged data would not do, as in
DVCS the GPD amplitude is added to the Bethe-Heitler amplitude with its
own spin structure, so using the reduced amplitudes would mean to obtain
the wrong interference terms in the expression for the cross section.

We realize that the bulk of the GPD
discussion\cite{ref.10-1,ref.10-2,ref.10-3} refers to a kinematics where
the transverse momentum of the virtual photon is not of order $Q$ but
small or zero (e.g. to the center-of-mass of virtual photon and target
hadron, or to the kinematics given by Eq.~(\ref{eq.05})).  Our concern
discussed in this work doesn't apply to this case and the contribution
of longitudinal photon polarization is indeed suppressed by 1/Q and can
be neglected in DVCS. We stress, however, that for a correct analysis of the
experimental data one must limit the reference frame to one where
the transverse momenta of the photons are small compared to Q. Such
frames exclude $q^+ = 0$, which is preferred in the case of form-factor
calculations.

Based on these straightforward tree-level calculations of DVCS amplitudes,
we conclude:

(i) The formulation of GPDs in the level of tensorial amplitude
${T_s}^{\mu\nu} + {T_u}^{\mu\nu}$ cannot be general enough to
cover the kinematics with large transverse momenta such as given by
Eqs.~(\ref{eq.04}) and (\ref{eq.06}) but is limited to the special
system of coordinates without involving large transverse momenta as
given by Eq.~(\ref{eq.05}).

(ii) In kinematics where the transverse components of the momenta are of
order $Q$ the full hadronic amplitudes and the reduced ones do not agree,
even in the limit $Q\to\infty$, which means that the calculations of the
DVCS amplitudes using the GPD cannot be trusted in this kinematics; In
addition, the contribution of the longitudinally polarized virtual photon
is not down by one order in $Q$ but even plays the role of cancelling
the singular parts;

(iii) The singularities we have found are in no way connected to the
strong-interaction part, but entirely due to the minus components of
the photon-polarization vectors, meaning that a calculation beyond tree
level will encounter the same singularities.

We have found~\cite{ref.11} the same singularities to occur in real Compton
scattering using the same kinematics. There they turn out to be of equal
magnitude but opposite sign in the $s$- and $u$-channel amplitudes and
thus cancel out, as expected.

This work is supported in part by the U.S. Department of Energy 
(No.DE-FG02-03ER41260).

\end{document}